\documentclass[a4paper,pre,aps,showpacs]{revtex4}

\usepackage{amsmath}
\usepackage{amsfonts,amssymb} 
\usepackage{times} 
\usepackage{graphicx}
\usepackage{float}
\usepackage{verbatim}
\usepackage{psfrag}
\newcommand{\aver}[1]{\langle #1 \rangle}
\newcommand{\wwp}{\frac{\omega_p^2}{\omega^2}}
\newcommand{\parvp}[1]{\frac{\partial #1}{\partial v_\perp}}
\newcommand{\parvl}[1]{\frac{\partial #1}{\partial v_\parallel}}
\allowdisplaybreaks[4]

\DeclareMathOperator{\real}{Re}

\newcommand{\jgr}{J. Geophys. Res.}
\newcommand{\grl}{Geophys. Res. Lett.}
\newlength{\mywidth}
\begin{document}
\title{Mirror modes: Nonmaxwellian distributions}
\author{M. Gedalin and Yu. E. Lyubarsky}
\affiliation{Ben-Gurion University, Beer-Sheva 84105, Israel}
\author{M. Balikhin}
\affiliation{ACSE, University of Sheffield, Mappin Street,
Sheffield S1 3JD, United Kingdom} 
\author{R.J. Strangeway and C.T.
Russell} 
\affiliation{IGPP/UCLA, 405 Hilgard Ave., Los Angeles, CA 90095-1567}
\date{\today}

\begin{abstract}
We perform direct analysis of mirror mode instabilities from the 
general dielectric 
tensor for several model distributions, in the longwavelength limit. 
The  growth rate  
at
the instability threshold depends on the derivative of the
distribution for zero parallel energy.  The maximum growth rate is
always $\sim k_\parallel v_{T\parallel}$ and the instability is of
nonresonant kind.  The instability growth rate and its dependence on
the propagation angle depend on the shape of the ion and electron
distribution functions.
\end{abstract}
\maketitle

\section{Introduction}\label{sec:intro}
Numerous observations of waves in the the Earth 
magnetosheath, as well as at other planets have stimulated studies of 
longwavelength and low-frequency modes in high $\beta$ magnetized 
plasmas. It has 
been theoretically shown that the features of low-frequency waves in 
hot plasmas differ significantly from those in cool plasmas, even in
the limit corresponding to the usual magnetohydrodynamic waves
\cite{Kr94}.  These findings have been subsequently proven by direct
comparison with observations \cite{Or94}.  However, particular
interest to the low-frequency modes in hot plasmas is explained by
observations of the mirror modes, which were found in planetary
magnetosheaths \cite{Ka70,Ts82,Vi95,Cz98}, in the solar wind
\cite{Wi94}, in cometary comas \cite{Ru87,Va89}, and in the wake of Io
\cite{Ki96,Ru99}.  These modes are nonpropagating zero frequency modes
(sometimes considered as the kinetic counterpart of the hydrodynamical
entropy mode), which are expected to grow in an anisotropic plasma
with sufficiently high $\beta_\perp/\beta_\parallel$ (see, e.g.,
\textcite{Ha75}).

Usual high amplitudes of observed 
mirror modes show that they easily achieve the nonlinear regime. At 
the same time, in several cases low-amplitude 
magnetic field structures with the same properties were observed 
which 
may
mean that the linear and nonlinear mirror mode features are 
generically related. Yet we do not know so far what makes these modes 
so ubiquitous and what determines their nonlinear amplitudes. 

The early explanation of the mirror instability \cite{Ha75} is based 
on the simple 
picture of the adiabatic response of the anisotropic pressure of 
magnetized particles. Numerical analyses of the mirror instability in 
bi-Maxwellian plasmas \cite{Ga92,MK92,Ga93} have shown that the 
maximum of the growth rate occurs at $k_\perp \rho_i\sim 1$ (where 
$\rho_i$ is the ion thermal gyroradius), which was  interpreted as an 
indication on the kinetic nature of the instability.

At the same time, \textcite{SK93} 
proposed a new explanation of the  instability mechanism as a 
resonant one, where the presence of a group of the resonant 
particles (with $v_\parallel=0$) plays the \textit{destructive} role 
in the mode excitation: the growth rate of the instability is claimed 
to be inversely proportional to the number of the resonant 
particles. This explanation was further reiterated with some 
modifications by \textcite{PS95} and \textcite{Po00}, and used by
\textcite{KS96} for the explanation of the nonlinear saturation
mechanism.  The analysis of \textcite{SK93} is done in the regime
where the phase velocity of the perturbation is much less than the
parallel thermal velocity, in other words, $\gamma \ll k_\parallel
v_{Ti\parallel}$, and therefore, is directly applied only at the very
threshold of the instability.  At the same time, numerical
calculations \cite{Ga93} show that most important events occur in the
range $\gamma \sim k_\parallel v_{Ti\parallel}$, which is not covered
in the previous analytical studies.

The previous analytical and numerical considerations of the linear regime
of the mirror instability, even in the longwavelength limit, are, as a
rule, restricted to the usage of the bi-Maxwellian distribution.  At
the same time particle distributions in collisionless plasma may
substantially differ from the Maxwellian.  For example, due to the ion
heating mechanism at the shock (see, e.g., \textcite{Sc90}), the
magnetosheath ion distributions may well deviate from the
bi-Maxwellian. It is therefore of interest to study the dependence of
the instability on the shape of the ion and electron distributions.

Yet another argument in favor of the analysis of other distributionsf is
that there is no good analytical approximations for the
dielectric tensor for the Maxwellian plasma in the range
$|\omega|/k_\parallel v_{T\parallel}\sim 1$, which forced researchers
to consider more convenient asymptotics.  It is, however, possible to
find the shapes of the distribution which allow closed analytical
presentation of the dielectric tensor in the whole range of phase
velocities and make the study of the instability physics more transparent. 

In the present paper we study in detail the dependence of the mirror
instability on the shape of the ion and electron distributions, using model
distribution functions which allow direct explicit analytical
calculation of the dielectric tensor.  We establish the generic
relation of the mirror instability with the oscillatory modes when the
Landau damping is absent and study the transition of damping modes to the
unstable regime.  We also propose an approximation which is useful for
the analytical treatment of the instability in the most important
range $\gamma \sim k_\parallel v_{Ti\parallel}$ in general case. 

The paper is organized as follows.  In section~\ref{sec:general} 
we 
derive the general dispersion relation in the longwavelength for 
arbitrary distribution 
function. In sections~\ref{sec:waterbag}-\ref{sec:hardbell} we apply 
the general analysis to three different distributions.  In
section~\ref{sec:threshold} we derive the instability condition and
the growth rate at the threshold for arbitrary distribution.  In
section~\ref{sec:bimax} we develop a useful approximation for the
analysis of the bi-Maxwellian-kind distributions  in the region of the
maximum growth rate.

\section{Dispersion relation in the longwavelength limit} 
\label{sec:general}

In what follows we will be interested in the longwavelength limit 
where $\omega\ll\Omega$ and $kv_T\ll\Omega$, while maintaining the
phase velocity finite $0<\omega/k<\infty$.  The last inequality means
that the phase velocity does not tend to zero in all propagation angle
range but it certainly may vanish for particular set of parameters. 
  For simplicity we  assume that both ions and
electrons are Maxwellian in the perpendicular direction, so that
$\aver{v_\perp^2}=2v_{T\perp}^2$ and $\aver{v_\perp^4}=8
v_{T\perp}^4$.  We also denote
$\aver{v_\parallel^2}=v_{T\parallel}^2$ and $\beta_{\parallel,\perp}
=2v_{T\parallel,\perp}^2\omega_p^2/c^2\Omega^2$ for each species
(subscript $i$ stands for ions and subscript $e$ for electrons).  Let us
 introduce the refraction index vector $\mathbf{N}=\mathbf{k}c/\omega$,
 such that $\mathbf{N}=(N_\perp, 0, N_\parallel) =N(\sin\theta, 0,
 \cos\theta)$.  With all this the components of the dispersion matrix
 $D_{ij}=N^2\delta_{ij}-N_iN_j -\epsilon_{ij}$ take the following form
 (see Appendix~\ref{sec:long}):
\begin{align}
& D_{11}=N_\parallel^2\left(1-\tfrac{1}{2}(\beta_\parallel 
-\beta_\perp)\right) -1 -\frac{\omega_{pi}^2}{\Omega_i^2}
-\frac{\omega_{pe}^2}{\Omega_e^2}, \label{eq:dismatrix11}\\
& D_{12}=0, \label{eq:dismatrix12}\\
& D_{13}=-N_\parallel N_\perp \left(1-\tfrac{1}{2}(\beta_\parallel 
-\beta_\perp)\right), \label{eq:dismatrix13}\\
& D_{22}= N^2\left( 1-\tfrac{1}{2} \cos^2\theta (\beta_\parallel - 
\beta_\perp) +\sin^2\theta \beta_\perp \right. \label{eq:dismatrix22}\\
&\left.-\sin^2\theta (r_i\beta_{i\perp 
}\bar{\chi}_i 
+r_e\beta_{e\perp }\bar{\chi}_e\right) -1 
-\frac{\omega_{pi}^2}{\Omega_i^2}-\frac{\omega_{pe}^2}{\Omega_e^2}, 
\notag\\
& D_{23}=-i\frac{\omega_{pi}^2\tan\theta}{\Omega_i\omega} 
\left(r_i\bar{\chi}_i - r_e\bar{\chi}_e\right), 
\label{eq:dismatrix23}\\
& D_{33}=N_\perp^2\left(1-\tfrac{1}{2}(\beta_\parallel 
-\beta_\perp)\right) 
-1 -\frac{\omega_{pi}^2\beta_{i\parallel}}{k_\parallel^2v_{Ti\parallel}^2} 
\left(\frac{\bar{\chi}_i}{\beta_{i\parallel}}
+ \frac{\bar{\chi}_e}{\beta_{e\parallel }} \right) \label{eq:dismatrix33}\\
& +\frac{\omega_{pi}^2\tan^2\theta }{\Omega_i^2}
r_i\bar{\chi}_i
+\frac{\omega_{pe}^2\tan^2\theta }{\Omega_e^2}
r_e\bar{\chi}_e, \notag
\end{align}
where $\beta_\parallel=\beta_{i\parallel}+\beta_{e\parallel}$, 
$\beta_\perp=\beta_{i\perp}+\beta_{e\perp}$, 
$r_i=\beta_{i\perp}/\beta_{i\parallel}$, 
$r_e=\beta_{e\perp}/\beta_{e\parallel}$,  and
\begin{equation}
\bar{\chi}=v_{T\parallel}^2\int 
(u-v_\parallel)^{-1}\parvl{f}dv_\parallel. \label{eq:barchi}
\end{equation}
The integration in \eqref{eq:barchi} is taken along the path below 
the singularity $v_\parallel=u$.  In what follows we shall also assume
that $\omega_{pi}^2/\Omega_i^2\gg 1$ and neglect unity relative to
this large parameter (which corresponds to the assumption $v_A\ll c$,
where $v_A=c\Omega_i/\omega_{pi}$ is the Alfven velocity).  In what
follows we also neglect
$\omega_{pe}^2/\Omega_e^2=(\omega_{pi}^2/\Omega_i^2)(m_e/m_i)$.  In
the above derivation we used
$\omega_{pi}^2/\Omega_i=-\omega_{pe}^2/\Omega_e$ in the quasineutral
electron proton plasma (this is not correct if any admixture of other
charged particles is present).

In the limit $\omega/\Omega_i\rightarrow 0$(and $\omega/k$ finite) 
the dispersion relation 
$D=\det \|D_{ij}\|=0$ splits into two ones.  One describes the purely
transverse Alfven wave (the wave electric field vector in the
$\mathbf{kB}_0$ plane, the wave magnetic field vector perpendicular to
the external magnetic field) with the dispersion
\begin{equation}
\omega_2=k^2v_A^2\cos^2\theta \left(1 -\tfrac{1}{2} 
(\beta_\parallel -\beta_\perp)\right). \label{eq:alfven}
\end{equation}
In this wave the absolute value of the magnetic field does not 
change, 
but the magnetic field rotates.

The second dispersion relation reads
\begin{equation}\label{eq:second}
\begin{split}
& \Psi(Z)=\left[2 - \cos^2\theta (\beta_\parallel -\beta_\perp) +
2\sin^2\theta \beta_\perp -2\sin^2\theta (
r_i\beta_{i\perp}\bar{\chi}_i +
r_e\beta_{e\perp}\bar{\chi}_e)\right.\\
&\left.
-Z^2\beta_{i\parallel}\cos^2\theta\right] \left[
\frac{\bar{\chi}_i}{\beta_{i\parallel}} + 
\frac{\bar{\chi}_e}{\beta_{e\parallel}}\right]
+\sin^2\theta\left[r_i \bar{\chi}_i -
r_e\bar{\chi}_e\right]^2=0,
\end{split}
\end{equation}
where we introduced $Z=\omega/k_\parallel v_{Ti\parallel}$  for 
convenience ($\omega$ 
is complex, in general, so that $Z=W+iG$), and 
$r_{i,e}=\beta_{i,e\perp}/\beta_{i,e\parallel}$.
Eq.~\eqref{eq:second} describes elliptically polarized waves with all three 
components of the wave electric field present, so that  in general there 
exists 
a nonzero component of the wave magnetic field  $B_z=N_\perp E_y $ in 
the direction of the external magnetic field. These waves not only 
rotate the magnetic field but change its magnitude as well. 

\begin{figure}[htb]
\psfrag{v}[][]{$v/v_{T\parallel}$}
\psfrag{f}[][]{$f(v_\parallel)$}
\includegraphics[width=\mywidth,keepaspectratio]{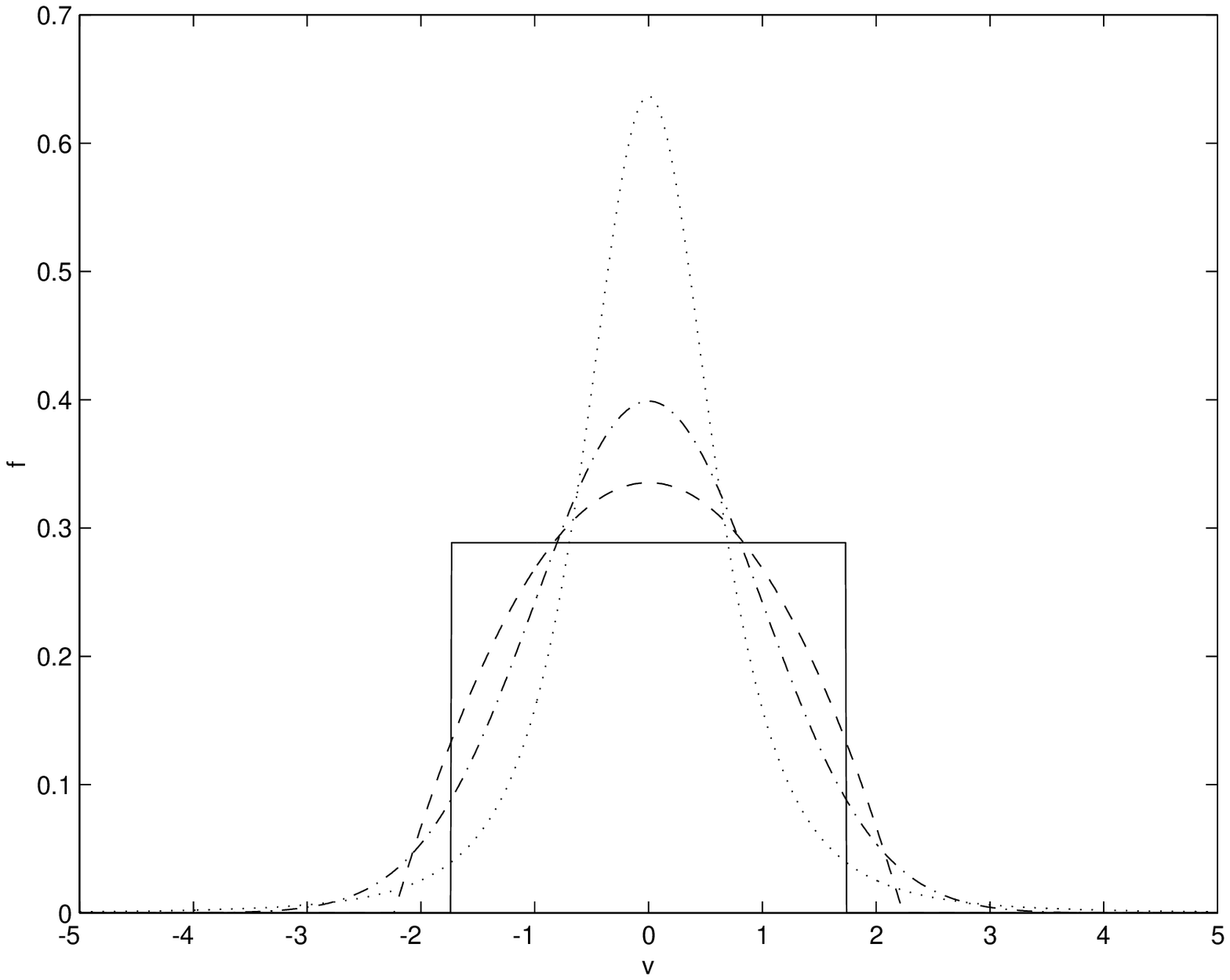}
\caption{Waterbag (solid line), hard bell (dashed), Lorentz (dotted), 
and Maxwellian (dash-dotted) distributions.}
\label{fig:functions}
\end{figure}

The functions $\bar{\chi}$ play the crucial role in the subsequent 
analysis.  They are  defined by the integral containing the 
distribution function $f(v_\parallel)$ and cannot be explicitly
calculated without particular choice of these distributions.  It is common
to choose $f$ as Maxwellian.  In this case $\bar{\chi}$ is
well-known and tabulated but has good asymptotic expansions only for
$|Z|\ll 1$ or $|Z|\gg 1$ (for electrons $Z$ should be substituted by
$Z(m_e/m_i)(v_{Ti\parallel}/v_{Te\parallel})$).  This actually
restricts possible analytical considerations of the mirror instability
only with the range $|Z|\ll 1$.  Yet, numerical analyses show that the
most important events occur in the vicinity of $|Z|\sim 1$ which is
unavailable to direct theoretical analysis when Maxwellian is chosen. 
On the other hand, there are vague indications that the qualitative
features of long waves (instabilities) in the high $\beta$ more or
less sensibly depend on the lowest moments of the distribution
function (provided it is sufficiently ``normal'': smooth, no beams, no
holes, etc.).  It therefore makes sense to investigate the dispersion
relations for a suitably chosen model distribution so that
$\bar{\chi}$ can be calculated and analyzed in the range $|Z|\sim 1$. 
In what follows we shall use three different distributions for these
purposes.  The waterbag distribution
$f=\Theta(v_0^2-v_\parallel^2)/2v_0$ will be used for study of the
behavior of longwavelength modes and their dependence on the plasma
parameters in the absence of Landau damping.  Here $\Theta(x)=1$ if
$x>0$ and $\Theta(x)=0$ if $x<0$.  The hard-bell distribution
$f=3(v_0^2-v_\parallel^2) \Theta (v_0^2-v_\parallel^2)/4v_0^3$ will
allow to include the Landau damping effects, and the Lorentz-like
distribution $f=(2v_0^3/\pi)(v_0^2+v_\parallel^2)^{-2}$ removes the
upper limit on the particle velocities.  The four distributions
(including Maxwellian $f=(2\pi v_{T\parallel}^2)^{-1/2}
\exp(-v_\parallel^2/2v_{T\parallel}^2)$) mentioned in this paper are
shown in Figure~\ref{fig:functions}.

\section{Waterbag}\label{sec:waterbag}
The waterbag distribution $f=\Theta(v_0^2-v_\parallel^2)/2v_0$ is
somewhat peculiar since the Landau damping is absent.  The analysis of
this distribution  allows to establish the generic relation of the
instability to nondamping propagating modes.   It is easy to find that in
this case
\begin{equation}
\bar{\chi}_i=\frac{1}{3-Z^2}, \qquad
\bar{\chi}_e=\frac{1}{3-Z^2\mu R}\label{eq:waterbagchi}
\end{equation}
where $\mu=m_e/m_i\approx 1/2000$,
$R=\beta_{i\parallel}/\beta_{e\parallel}$, and 
$v_{T\parallel}^2=v_0^2/3$.  In the limit $Z=0$ one has
$d\equiv \bar{\chi}(Z=0)=1/3$.  It is worth noting that for the Maxwellian
distribution $d=1$.  In this section we use for electrons the
approximation of the massless bi-Maxwellian (instead of above
waterbag,which is used only for ions), for which $\bar{\chi}_e=1$.  The
resulting dispersion relation \eqref{eq:second} is a third order
equation with respect to $Z^2$ with real coefficients.  Although this
equation can be analyzed directly and even solved analytically,
graphical representation of the roots is much more convenient.

\begin{figure}[htb]
\psfrag{theta}[][]{$\theta$}
\psfrag{omega}[][]{$\omega/k_\parallel v_{Ti\parallel}$}
\includegraphics[width=\mywidth,keepaspectratio]{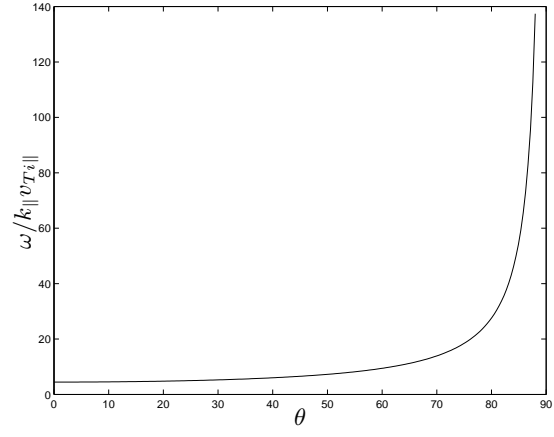}
\caption{Phase velocity of the fast mode as a function of propagation 
angle for the case of the waterbag distribution with 
$\beta_{i\parallel}=\beta_{i\perp}=\beta_{e\parallel}=\beta_{e\perp}=0.1$
and massless bi-Maxwellian electrons.}
\label{fig:waterbag01f}
\end{figure}

\begin{figure}[htb]
\psfrag{theta}[][]{$\theta$}
\psfrag{omega}[][]{$\omega/k_\parallel v_{Ti\parallel}$}
\includegraphics[width=\mywidth,keepaspectratio]{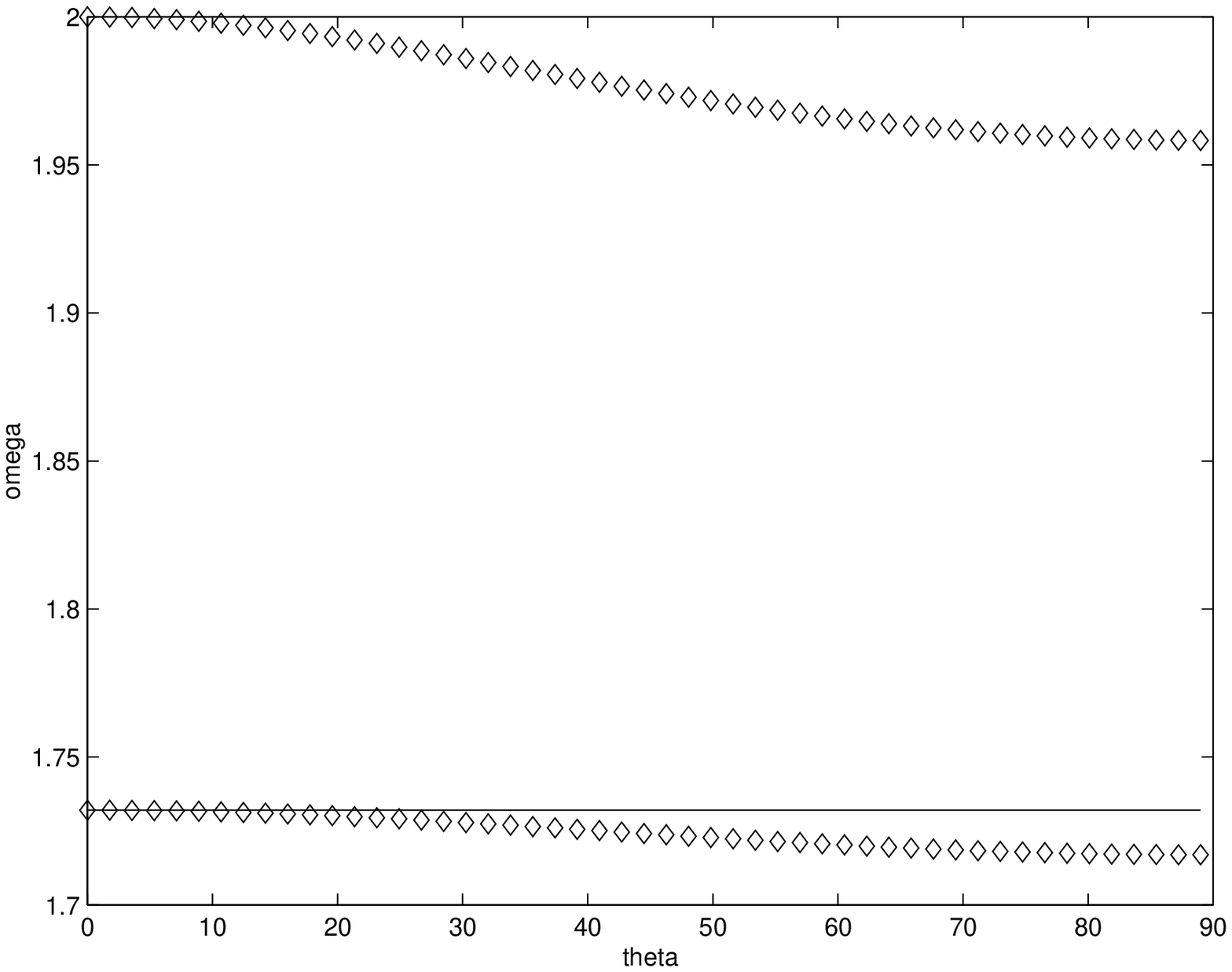}
\caption{Phase velocity (diamonds) 
of the two low-velocity modes as a function of propagation 
angle for the case of the waterbag distribution with 
$\beta_{i\parallel}=\beta_{i\perp}=\beta_{e\parallel}=\beta_{e\perp}=0.1$
and massless bi-Maxwellian electrons.  The solid line is $\omega=\sqrt{a_i}
k_\parallel v_{Ti\parallel}=k_\parallel v_{0i}$.  }
\label{fig:waterbag01low}
\end{figure}

Figure~\ref{fig:waterbag01f}
shows the mode with the highest phase velocity (fast mode) for the 
case when $\beta_{i\parallel}=\beta_{i\perp}= 
\beta_{e\parallel}=\beta_{e\perp}=0.1$ and massless bi-Maxwellian
electrons.  It it worth noting that ions are \emph{not} isotropic
since they are Maxwellian in the perpendicular direction and waterbag
in the parallel direction.  The phase velocity of the fast mode is
well above $k_\parallel v_{Ti\parallel}$ so that it has nothing to do
with the mirror instability.  We do not consider this mode in the rest
of the paper.  We do not consider the Alfven mode either.  The
remaining two low-phase velocity modes are shown in
Figure~\ref{fig:waterbag01low} together with $\omega=k_\parallel
v_{0i}$ (solid line).  The upper curve is above the resonant region
having $|\omega|>|k_\parallel v_\parallel|$ for all ions.  The lower
mode is inside the resonant region and would damp if there were
nonzero $\partial f/\partial v_\parallel$.

\begin{figure}[htb]
\psfrag{theta}[][]{$\theta$}
\psfrag{omega}[][]{$\omega/k_\parallel v_{Ti\parallel}$}
\includegraphics[width=\mywidth,keepaspectratio]{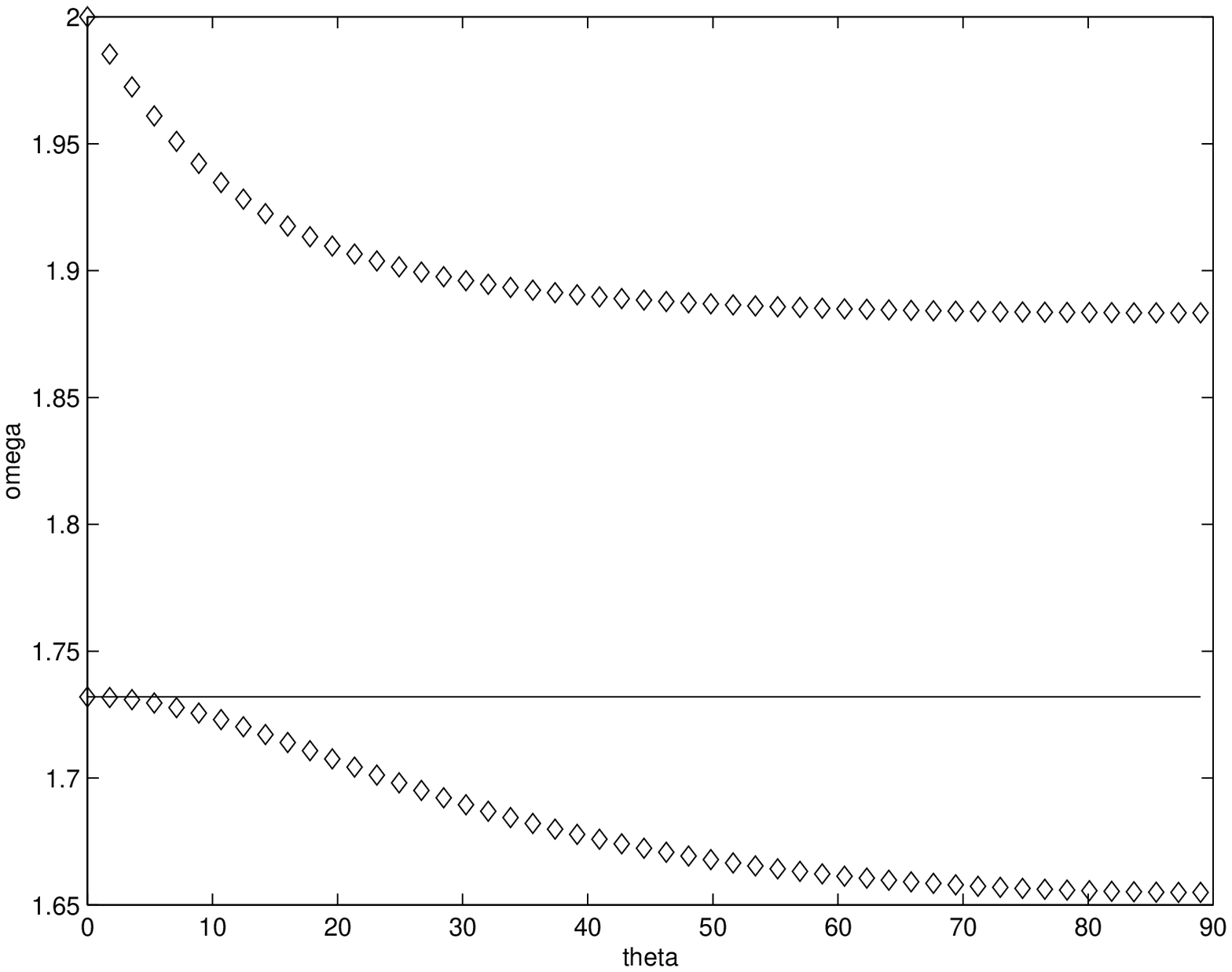}
\caption{Phase velocity (diamonds) 
of the two low-velocity modes as a function of propagation 
angle for the case of the waterbag distribution with 
$\beta_{i\parallel}=\beta_{i\perp}=\beta_{e\parallel}=\beta_{e\perp}=0.5$
and massless bi-Maxwellian electrons.  The solid line is
$\omega=\sqrt{a_i} k_\parallel v_{Ti\parallel}=k_\parallel v_{0i}$.  }
\label{fig:waterbag05low}
\end{figure}

Figure~\ref{fig:waterbag05low}
shows the same two modes but in the case 
$\beta_{i\parallel}=\beta_{i\perp}=\beta_{e\parallel}=\beta_{e\perp}=0.5$. 
In both cases the naive instability condition 
\begin{equation}
K=\beta_\perp/\beta_\parallel - 2 - 2/\beta_\perp>0 \label{eq:naive}
\end{equation}
is not fulfilled, although in the second case $K$ is
closer to the threshold just because of the larger $\beta_\perp$. 
There is no much difference in the behavior of the two modes for these
two cases, except a little stronger decrease of the phase velocities
towards the perpendicular propagation regime in the higher
$\beta_\perp$ case.

\begin{figure}[htb]
\psfrag{theta}[][]{$\theta$}
\psfrag{wg}[][]{$\omega/k_\parallel v_{Ti\parallel}$, 
$\gamma/k_\parallel v_{Ti\parallel}$}
\psfrag{w}[][]{frequency $W$}
\psfrag{g}[][]{growth rate $G$}
\includegraphics[width=\mywidth,keepaspectratio]{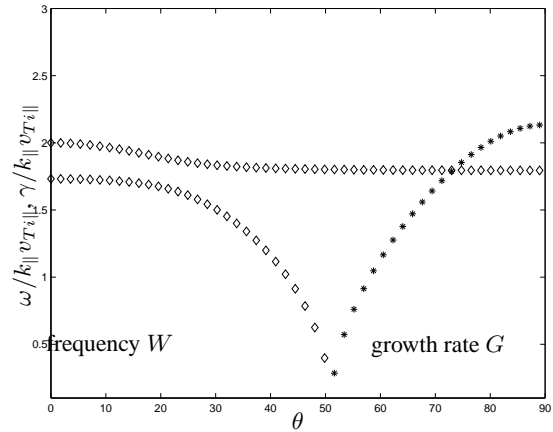}
\caption{Behavior 
of the two low-velocity modes as a function of propagation 
angle for the case of the waterbag distribution with 
$\beta_{i\parallel}=\beta_{e\parallel}=0.1$, 
$\beta_{i\perp}=\beta_{e\perp}=0.5$ (so that $K=1$), 
and massless bi-Maxwellian electrons. Diamonds mark the modes in the range 
where their frequencies are purely real, stars show the growth rate 
of the aperiodic instability.  }
\label{fig:waterbag15em}
\end{figure}

Figure~\ref{fig:waterbag15em}
shows the behavior of the two modes in the anisotropic case
$\beta_{i\parallel}=\beta_{e\parallel}=0.1$, 
$\beta_{i\perp}=\beta_{e\perp}=0.5$ (so that $K=1$), 
and bi-Maxwellian electrons. The lower mode now remains purely 
propagating mode for smaller angles (diamonds) but turns into an 
aperiodic instability for larger angles of propagation (stars). The 
obvious conclusion from Figure~\ref{fig:waterbag15em} is that the 
unstable mode has its propagation counterpart for the smaller angles 
of propagation. The relative growth rate $G=\gamma/k_\parallel 
v_{Ti\parallel} \sim 1$ is large in the whole range of instability, 
so 
that the approximation $G\ll 1$ \cite{SK93} is not applicable. 

\begin{figure}[htb]
\psfrag{theta}[][]{$\theta$}
\psfrag{wg}[][]{$\omega/k_\parallel v_{Ti\parallel}$}
\psfrag{w}[][]{frequency $W$}
\psfrag{g}[][]{growth rate $G$}
\includegraphics[width=\mywidth,keepaspectratio]{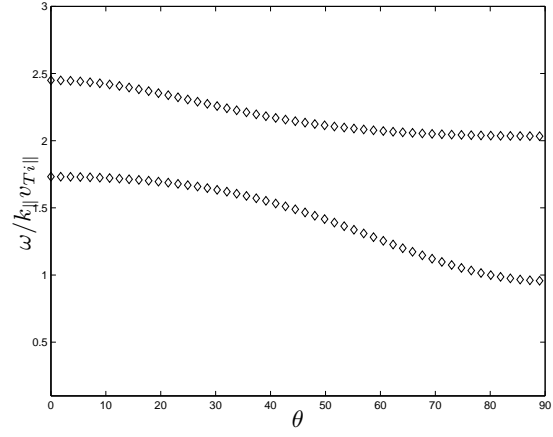}
\caption{Behavior 
of the two low-velocity modes as a function of propagation 
angle for the case of the waterbag distribution with 
$\beta_{i\parallel}=\beta_{e\parallel}=0.1$, 
$\beta_{i\perp}=\beta_{e\perp}=0.5$ (so that $K=1$), 
and waterbag electrons, $d_e=1/3$. There is no instability.  }
\label{fig:waterbag15ew}
\end{figure}

It is of interest to compare this case with the massless waterbag
electrons $\bar{\chi}_e=1/3$.  The corresponding curves in
Figure~\ref{fig:waterbag15ew} show that there is no instability in
this case despite the fact that the threshold \eqref{eq:naive} is exceeded.

Thus, the analysis of the waterbag distribution already shows that 
(a) there 
is, in general, the propagating counterpart of the mirror 
instability if Landau damping is absent, (b) the instability threshold and
growth rate are sensitive to the details of the distribution and not
only to the second moment, and (c) the instability is aperiodic, that
is, in the unstable range $W=0$ and $G>0$.  It can be shown that the
last feature is generally valid unless the distribution function is
very peculiar (see Appendix~\ref{sec:aperiodic}).

\section{Hard-bell and Lorentz distributions}\label{sec:hardbell}
The waterbag distribution does not allow Landau damping since 
$\partial f/\partial v_\parallel =0$ everywhere. In order to get rid 
of this restriction we consider the hard-bell distribution 
$f=3(v_0^2-v_\parallel^2) \Theta (v_0^2-v_\parallel^2)/4v_0^3$, which
has nonzero derivative but is is compact ($f=0$ for
$|v_\parallel|>v_0$ .  In this case
\begin{equation}\label{eq:hardbellchi}
\begin{split}
& \bar{\chi}_i=\frac{3}{5} \left[1+\frac{Z}{4\sqrt{5}} 
\ln\frac{(\sqrt{5} -W)^2 +G^2}{(\sqrt{5} +W)^2 +G^2}\right.\\
& \left.+\frac{iZ}{2\sqrt{5}}\left(\arctan\frac{\sqrt{5}-W}{G} 
+ \arctan\frac{\sqrt{5}+W}{G}\right)\right], 
\end{split}
\end{equation}
where $Z=W+iG$, $W$ and $G$ being real,  $G>0$, and 
$v_0^2=5v_{T\parallel}^2$.  The corresponding $d=\bar{\chi}(Z=0)=3/5$. 
The corresponding expression for $\bar{\chi}_e$ is obtained from
\eqref{eq:hardbellchi} by substitution $Z\rightarrow Z\sqrt{\mu R}$. 

In order to analyze non-compact distributions too we shall consider 
the Lorentz distribution 
$f=(2v_0^3/\pi)(v_0^2+v_\parallel^2)^{-2}$.  In this case
\begin{equation}
\bar{\chi}_i=\frac{16iZ}{(1+Z^2)^3}+\frac{3i}{i-Z}
-\frac{2Z}{(i-Z)^3} +\frac{3iZ}{(i-Z)^2}, \label{eq:lorentzchi}
\end{equation}
with $v_0^2=v_{T\parallel}^2$ and $d=3$.  Again, 
$\bar{\chi}_e$ is obtained  by substitution $Z\rightarrow Z\sqrt{\mu R}$.

We shall also compare the results for these distributions with the
bi-Maxwellian.  In this case there is no compact analytical
expression for $\chi$ and we use direct numerical calculation. 

In what follows we are interested only in the unstable region. The 
subparticle mode is expected to be strongly damped in the propagation
range.  The ``superparticle'' mode is not damped in the hard-bell case
and almost not damped in the Lorentz case.

\begin{figure}[htb]
\psfrag{theta}[][]{$\theta$}
\psfrag{g}[][]{$\gamma/k_\parallel v_{Ti\parallel}$}
\psfrag{gamma}[][]{$\gamma/kv_{Ti\parallel}$}
\includegraphics[width=\mywidth,keepaspectratio]{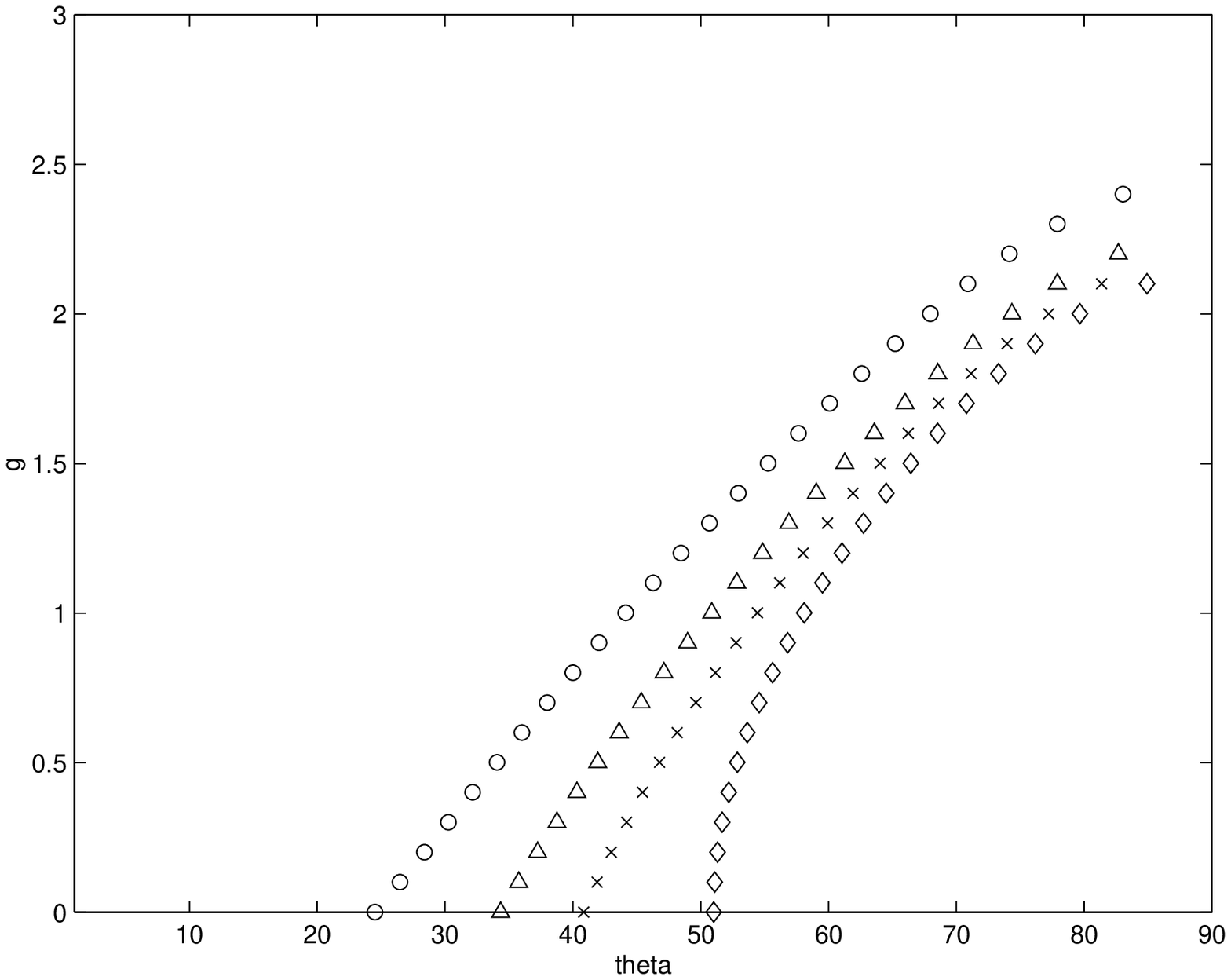}
\caption{Growth rates for the mirror instability in the case  
of $\beta_{i\parallel}=\beta_{e\parallel}=0.1$, 
$\beta_{i\perp}=\beta_{e\perp}=0.5$, and massless bi-Maxwellian electrons 
$d_e=1$, and four different distributions: waterbag (diamonds), 
hard-bell (crosses), Lorentz (circles), and bi-Maxwellian (triangles). }
\label{fig:four15em}
\end{figure}

\begin{figure}[htb]
\psfrag{theta}[][]{$\theta$}
\psfrag{g}[][]{$\gamma/k_\parallel v_{Ti\parallel}$}
\psfrag{gamma}[][]{$\gamma/kv_{Ti\parallel}$}
\includegraphics[width=\mywidth,keepaspectratio]{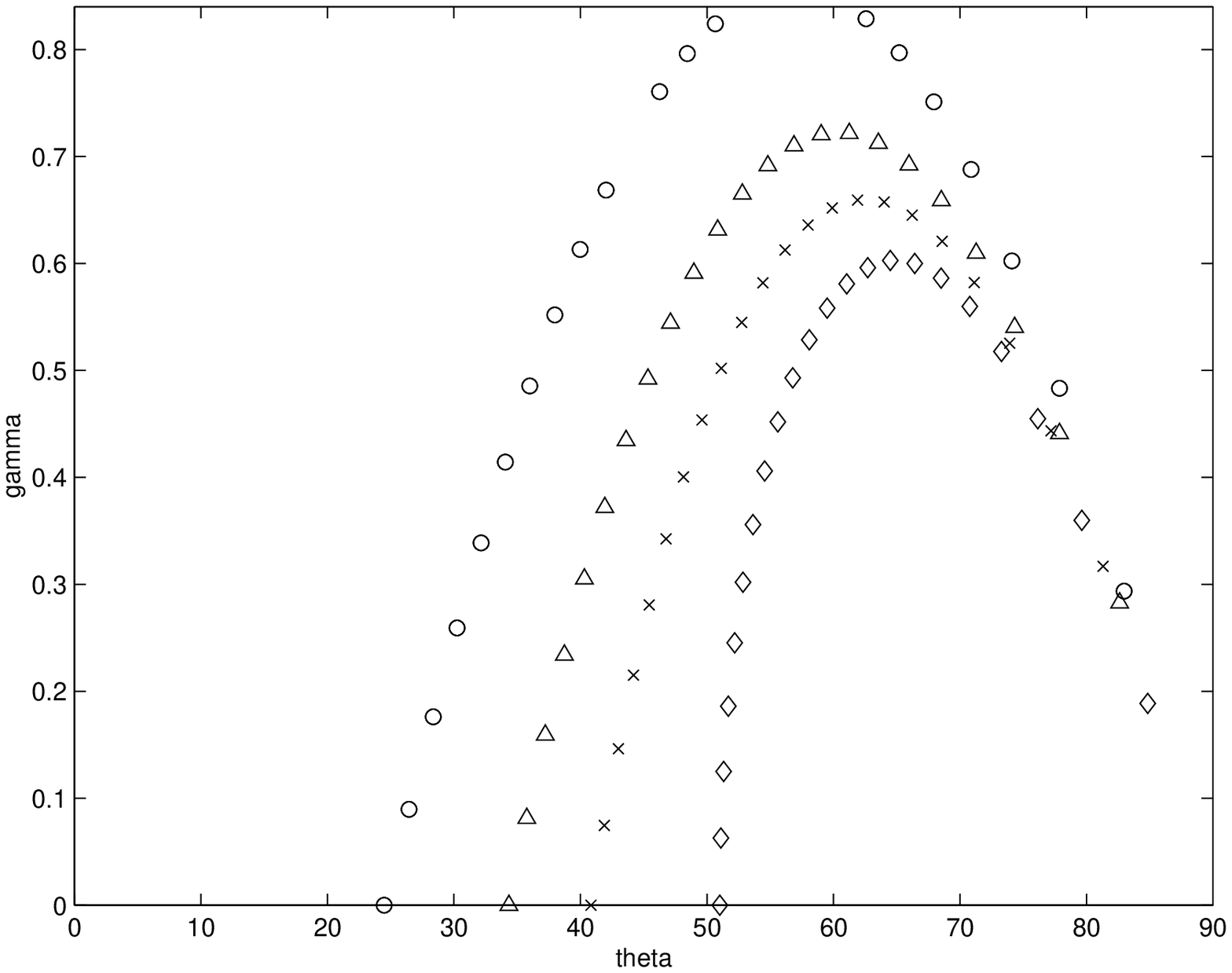}
\caption{Growth rates for the mirror instability in the case  
of $\beta_{i\parallel}=\beta_{e\parallel}=0.1$, 
$\beta_{i\perp}=\beta_{e\perp}=0.5$, and massless bi-Maxwellian electrons 
$d_e=1$, and four different distributions: waterbag (diamonds), 
hard-bell (crosses), Lorentz (circles), and bi-Maxwellian (triangles),
normalized on $kv_{Ti\parallel}$.  }
\label{fig:fourg1}
\end{figure}

As the first set of parameters for the unstable regime we choose
$\beta_{i\parallel}=\beta_{e\parallel}=0.1$,
$\beta_{i\perp}=\beta_{e\perp}=0.5$, and massless bi-Maxwellian electrons
$\bar{\chi}_e=1$.  Figure~\ref{fig:four15em} shows the growth rates for
the three distributions.  The highest growth rate is for the
Lorentzian, the lowest is for the waterbag.  Figure~\ref{fig:fourg1}
shows the same growth rates as in Figure~\ref{fig:four15em} but
normalized on $kv_{Ti\parallel}$ which allows to compare growth rates
of the modes with the same wavenumber $k$ and different angles of
propagation.  It is seen that the maximum growth rates is achieved
approximately at the same angle of propagation $\approx 60^\circ$ for
all distributions, but the threshold angle moves towards more
quasiparallel regimes for distributions with stronger tails
(Maxwellian and Lorentzian).

\begin{figure}[htb]
\psfrag{theta}[][]{$\theta$}
\psfrag{g}[][]{$\gamma/k_\parallel v_{Ti\parallel}$}
\psfrag{gamma}[][]{$\gamma/kv_{Ti\parallel}$}
\includegraphics[width=\mywidth,keepaspectratio]{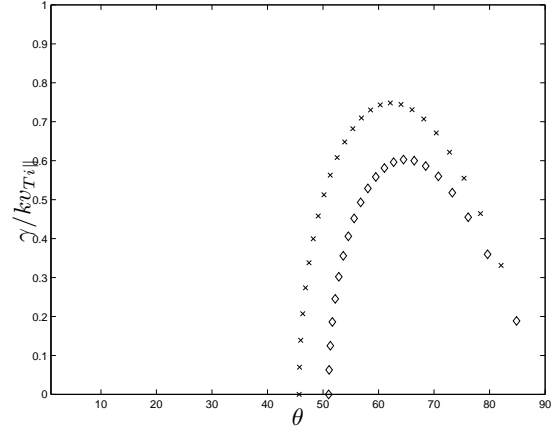}
\caption{Dependence of the growth rate on $\beta_\perp$ with
$K=1$ for waterbag ions and massless bi-Maxwellian
electrons: diamonds correspond to 
$\beta_{i\parallel}=\beta_{e\parallel}=0.1$,
$\beta_{i\perp}=\beta_{e\perp}=0.5$, crosses correspond to 
$\beta_{i\parallel}=\beta_{e\parallel}=0.25$,
$\beta_{i\perp}=\beta_{e\perp}=1$.  }
\label{fig:wai125}
\end{figure}

Figure~\ref{fig:wai125} shows the dependence of the growth rate on
$\beta_\perp$ when $K=1$ and $\beta_{i\perp}/\beta_{e\parallel}=
\beta_{i\parallel}/\beta_{e\parallel}=1$ remain constant.  Both
curves correspond to the waterbag ions and massless bi-Maxwellian
electrons.  Diamonds stand for the same parameters as in
Figure~\ref{fig:waterbag15em}, crosses correspond to
$\beta_{i\perp}=1$ and $\beta_{i\parallel}=0.25$.  The instability is
stronger for higher $\beta_\perp$. 

In the previous analysis we always used the approximation of massless
bi-Maxwellian distribution corresponding to $\chi_e=1$. 
Figures~\ref{fig:threeown} and~\ref{fig:threeg1own} show the growth
rate of the instability when the electron distributions are chosen in
the same form as the ion distributions.  One can see that the
waterbag distributions become stable, while the growth rate in the
case of Lorentzian drastically increases.  The ratio of the maximum
growth rates shown in Figures~\ref{fig:fourg1}
and~\ref{fig:threeg1own} roughly corresponds to $d_e=\chi_e(Z=0_)$
which shows that the maximum growth rate significantly
on electrons (see sections~\ref{sec:threshold} and~\ref{sec:bimax}).

\begin{figure}[htb]
\psfrag{theta}[][]{$\theta$}
\psfrag{g}[][]{$\gamma/k_\parallel v_{Ti\parallel}$}
\psfrag{gamma}[][]{$\gamma/kv_{Ti\parallel}$}
\includegraphics[width=\mywidth,keepaspectratio]{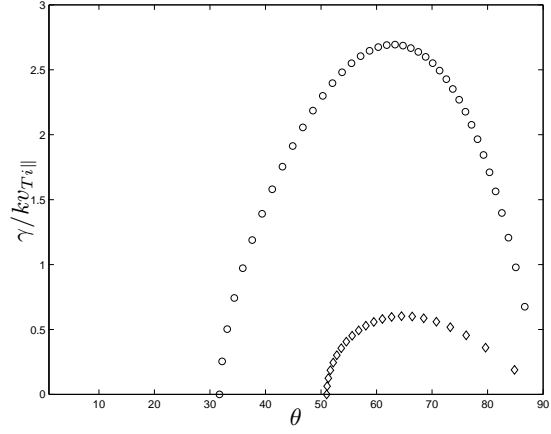}
\caption{Growth rates for the mirror instability in the
case 
of $\beta_{i\parallel}=\beta_{e\parallel}=0.1$, 
$\beta_{i\perp}=\beta_{e\perp}=0.5$, waterbag ions and two different
electron distributions: massive bi-Maxwellian (diamonds) and Lorentz
(circles).  }
\label{fig:waimaeloe}
\end{figure}

For other combinations of ion and electron distributions the ratios
may be even greater as is seen in Figure~\ref{fig:waimaeloe}, where
diamonds correspond to waterbag ions and massless bi-Maxwellian
electrons, while circles correspond to waterbag ions and Lorentz
electrons.  The $\beta$ parameters are the same for both cases.

\section{Near the threshold}\label{sec:threshold}
It is possible to obtain general results just above the threshold of the 
instability, where $Z=iG\rightarrow +0$. For $f=f(v^2)$ it is easy to find 
\begin{equation}\label{eq:chith}
\begin{split}
& \chi=\int \frac{1}{iG-v_\parallel}\parvl{f}dv_\parallel\\
&=-\int \frac{v_\parallel}{v_\parallel^2+G^2}\parvl{f}dv_\parallel
=-\int \frac{df}{d\mathcal{E}}dv_\parallel +G\int \frac{G}
{v_\parallel^2+G^2}\frac{df}{d\mathcal{E}}dv_\parallel\\
&=-\int \frac{df}{d\mathcal{E}}dv_\parallel +\pi G 
\frac{df}{d\mathcal{E}}|_{v_\parallel=0}=
d-\kappa G, 
\end{split}
\end{equation}
where $\mathcal{E}=v_\parallel^2/2$ is the energy (on the unit mass). 
Substituting this into \eqref{eq:second} and neglecting all terms of 
the order $Z^2$ and higher, one has
\begin{align}
& G=-A/B, \label{eq:G}\\
& A=[2-\cos^2\theta(\beta_\parallel-\beta_\perp) + 2\sin^2\theta 
\beta_\perp -2\sin^2\theta(r_i\beta_{i\perp}d_i\notag\\
&+r_e\beta_{e\perp}d_e)]\left(\frac{d_i}{\beta_{i\parallel}}
+ \frac{d_e}{\beta_{e\parallel}}\right) + \sin^2\theta 
(r_id_i-r_ed_e)^2, \label{eq:A}\\
& B=-\frac{\kappa_i}{\beta_{i\parallel}} [
2-\cos^2\theta(\beta_\parallel-\beta_\perp) + 2\sin^2\theta 
\beta_\perp -2\sin^2\theta(r_i\beta_{i\perp}d_i\notag\\
& +r_e\beta_{e\perp}d_e)] +2\sin^2\theta r_i\beta_{i\perp} \kappa_i 
\left(\frac{d_i}{\beta_{i\parallel}}
+ \frac{d_e}{\beta_{e\parallel}}\right)  -2\sin^2\theta 
\kappa_i(r_id_i-r_ed_e), \label{eq:B}
\end{align}
where we neglected $\kappa_e\sim \kappa_i\sqrt{m_e/m_i}$. 
The instability threshold for given $\theta$ is found from the 
condition $G=0$, that is, $A=0$, which gives
\begin{equation}\label{eq:condition1}
\begin{split}
& 2+\beta_\perp-\beta_\parallel +\sin^2\theta [\beta_\parallel 
+\beta_\perp -2(r_i\beta_{i\perp}d_i +r_e\beta_{e\perp}d_e)\\
& + (r_id_i-r_ed_e)^2/(d_i/\beta_{i\parallel} + 
d_e/\beta_{e\parallel})]=0.
\end{split}
\end{equation}
Since $0\leq\sin^2\theta\leq1$, the global instability criterion 
reads (in the assumption that $2+\beta_\perp>\beta_\parallel$):
\begin{equation}
2(r_i\beta_{i\perp}d_i +r_e\beta_{e\perp}d_e)-2 -2\beta_\perp -
(r_id_i-r_ed_e)^2\left(\frac{d_i}{\beta_{i\parallel}}
+ \frac{d_e}{\beta_{e\parallel}}\right)^{-1}>0.\label{eq:condition2}
\end{equation}

It is instructive to consider several simple cases. 
We can neglect completely the electron contribution  by putting $d_e=0$,
which gives the instability criterion in the form
\begin{equation}
\frac{d\beta_\perp}{\beta_\parallel}>2(1+\frac{1}{\beta_\perp}), 
\label{eq:condition3}
\end{equation}
and for the bi-Maxwellian distribution, $d=1$, reduces to the naive
mirror instability criterion.

On the other hand, when $r_e=r_i=\beta_\perp/\beta_\parallel$ and
$d_e=d_i=d$, one gets 
\begin{equation}
\frac{d\beta_\perp}{\beta_\parallel}>1+\frac{1}{\beta_\perp}.
\label{eq:condition4}
\end{equation}
This condition is harder for more compact distributions ($d=1/3$ for
waterbag and $d=3/5$ for hard-bell) and softer for distributions with
long tails ($d=1$ for Maxwellian and $d=3$ for Lorentzian).
The global instability condition \eqref{eq:condition2} can be written
in a more symmetric form as follows:
\begin{equation}\label{eq:condition5}
\begin{split}
& (r_id_i+r_ed_e)^2 +
\frac{2(\beta_{i\perp}^2+\beta_{e\perp}^2) d_ed_i}{
\beta_{i\parallel}\beta_{e\parallel}} \\
& -2(1+\beta_\perp)\left(\frac{d_e}{\beta_{e\parallel}}
+\frac{d_i}{\beta_{i\parallel}}\right)>0
\end{split}
\end{equation}
which emphasizes the symmetric role of ions and electrons in the
instability onset (cf.  \textcite{Po00}).

Indeed, near the threshold $\gamma/k_\parallel
v_{T\parallel}\ll 1$ and the response of both electrons and ions is
adiabatic, that is, their inertia does not play any role.  In these
circumstances the mass of the particle is of not importance.  Their
role in the response to the parallel electric field is, however,
antisymmetric because of the different signs of the charge: the
adiabatic response is obtained from
$eE_z-(1/n)(dp/dz)=eE_z-ik_\parallel p/n=0$.  The parallel response plays the
crucial role in the instability development.  As is known the
instability occurs because of the breakdown of the local frozen-in
condition and efficient drag of particles out of the field enhancement
into the field depletion region \cite{SK93,PS95,Po00}.  Thus, when the
magnetic field is perturbed, $B_z=B_0+\delta B_z$, the perturbation of
the density of the species $s$ is
\begin{equation}
\frac{\delta n_s}{n_{0s}}=\frac{\delta B_z}{B_0} +
\frac{\delta n_s^{\text{(ext)}}}{n_{0s}},
\end{equation}
where $\delta
n_s^{\text{(ext)}}$ is due to the motion along the field lines.  In the
adiabatic regime $\gamma/k_\parallel v_{T\parallel}\ll 1$ this change can
be considered as a quasistatic response to the effective potential
$\phi_{\text{eff}}=\phi+\mu_s \delta B_z/q_s$, where $\phi$ is the
electrostatic potential, $\mu_s=\aver{v_\perp^2}_s/2B_0 $ is the average
magnetic moment, and $q_s$ is the charge of the species.  The density
response to this effective potential can be found from the reduced
Vlasov equation
\begin{equation}
\frac{\partial f_s}{\partial t} + v_\parallel \frac{\partial f_s}{\partial z} = 
q_s \frac{\partial \phi_{\text{eff}}}{\partial z} \parvl{f_s}, 
\end{equation}
which for $\partial/\partial t=\gamma$ and $\partial /\partial
z=ik_\parallel$ gives
\begin{equation}
\frac{\delta n_s^{\text{(ext)}}}{n_{0s}}=\phi_{\text{eff}} \int
\frac{ik_\parallel}{\gamma + ik_\parallel v_\parallel}
\parvl{f_s}dv_\parallel.\label{eq:deltan}
\end{equation}
It is easy to see that in the adiabatic regime near
the threshold of the instability, $\gamma\rightarrow 0$, this expression 
reduces to the following
\begin{equation}
\frac{\delta
n_s^{\text{(ext)}}}{n_{0s}}=-\frac{q_s\phi_{\text{eff}}}{4\pi n_{0s}q_s^2
r_D^2},
\end{equation}
where $r_D$ is the Debye length calculated with the parallel
distribution function.  It is easy to see that
$r_D^2=v_{T\parallel}^2/\omega_p^2 d$, where $d=\bar{\chi}(Z=0)$. The
electrostatic potential $\phi$ can be excluded using the
quasineutrality condition $\delta n_e=\delta n_i$, which eventually
gives
\begin{equation}
\frac{\delta n}{n}=\frac{\delta B_z}{B_0} \left[ 1 - \frac{T_{e\perp}
+T_{i\perp}}{4\pi e^2n_0(r_{De}^2+r_{Di}^2)}\right],\label{eq:debye}
\end{equation}
where we have taken into account that $\mu=T_\perp/B_0$. 
Eq.~\eqref{eq:debye} shows that smaller Debye lengths $r_D$ (larger
$d$) result in the stronger drag of the particles into the weak field
region, that reducing the kinetic pressure response to the magnetic
field enhancement and supporting instability.  Therefore, stronger
Debye screening (larger $d$) would lower the instability threshold, in
agreement with the found from rigorous calculations.

From \eqref{eq:G}--\eqref{eq:B} it is easily seen that the growth 
rate is inversely proportional to 
$\kappa_i=-\pi(df/d\mathcal{E})|_{v_\parallel=0}$, and not to the 
number of particles with $v_\parallel=0$ (cf. \textcite{SK93}).  The
latter is correct for the bi-Maxwellian distribution since $
(df/d\mathcal{E})\propto f$ in this case.  For other distributions
this relation may well be wrong.  For example, for the waterbag
distribution $(df/d\mathcal{E})|_{v_\parallel=0}=0$ and higher order
terms should be retained to investigate the behavior near the
threshold.  It is easy to see from~\eqref{eq:chith} that in this case
$\bar{\chi}=d-\alpha G^2$, where $\alpha=-\int v_\parallel^{-2}
(df/d\mathcal{E}) dv_\parallel$ is well-defined.  The dispersion
relation~\eqref{eq:second} becomes than a first order equation for
$G^2$, which has one positive solution near the threshold.  It is
clear that in this case the growth rate is determined by the whole
distribution and not only by the behavior in $v_\parallel=0$.

\section{Hydrodynamical regime}
\label{sec:bimax}
\begin{figure}[htb]
\psfrag{theta}[][]{$\theta$}
\psfrag{gamma}[][]{$\gamma/k_\parallel v_{Ti\parallel}$}
\includegraphics[width=\mywidth,keepaspectratio]{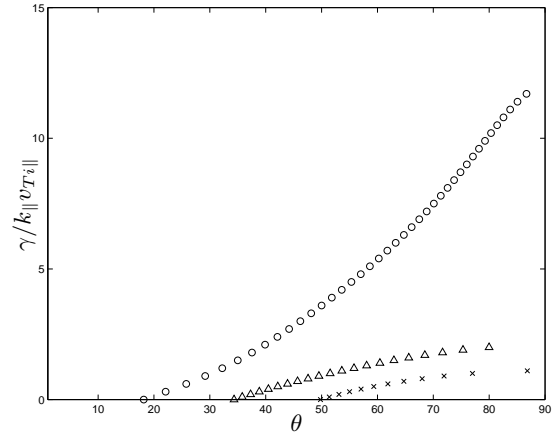}
\caption{Growth rates for the mirror instability in the case  
of $\beta_{i\parallel}=\beta_{e\parallel}=0.1$, 
$\beta_{i\perp}=\beta_{e\perp}=0.5$, and three different combinations:
 hard-bell ions and electrons (crosses), Lorentz ions and
 electrons (circles), and bi-Maxwellian ions and (massive) electrons
 (triangles). }
\label{fig:threeown}
\end{figure}

\begin{figure}[htb]
\psfrag{theta}[][]{$\theta$}
\psfrag{g}[][]{$\gamma/k_\parallel v_{Ti\parallel}$}
\psfrag{gamma}[][]{$\gamma/kv_{Ti\parallel}$}
\includegraphics[width=\mywidth,keepaspectratio]{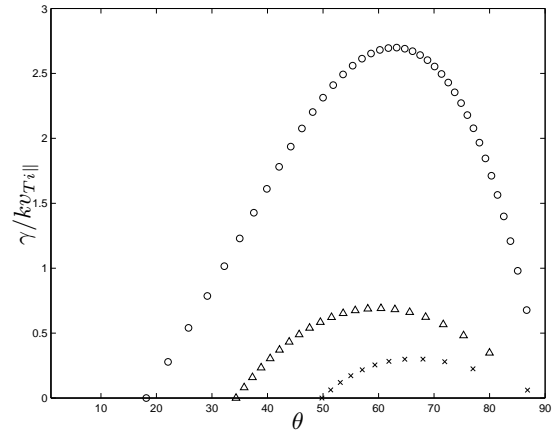}
\caption{Same as in Figure~\protect\ref{fig:threeown} but normalized
on $kv_{Ti\parallel}$. }
\label{fig:threeg1own}
\end{figure}

The previous analysis shows that maximum $Z$ is always of the order
of unity or larger, which means that ions no longer respond
adiabatically to the magnetic field enhancements and their inertia
begins to play an important role.  This also means that it is thermal
particles of the ion distribution body with $v\sim v_{Ti\parallel}$
which are mainly responsible for the instability development and not
the group of resonant particles with $v_\parallel=0$. 
Figure~\ref{fig:threeown} shows that for some distributions the
instability may be very fast so that the electron inertia should be
taken into account.

\begin{figure}[htb]
\psfrag{g}[][]{$G$}
\psfrag{chi}[][]{$\chi$}
\includegraphics[width=\mywidth,keepaspectratio]{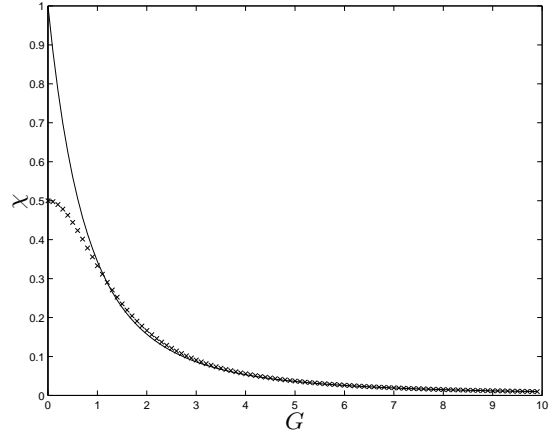}
\caption{Approximation of $\chi(G)$ for the Maxwellian distribution. 
The numerically calculated $\chi(G)$ (solid line) is compared to  $
\chi=(G^2+2)^{-1}$ (crosses).}
\label{fig:compchi}
\end{figure}

\begin{figure}[htb]
\psfrag{g}[][]{$G$}
\psfrag{chi}[][]{$\chi$}
\includegraphics[width=\mywidth,keepaspectratio]{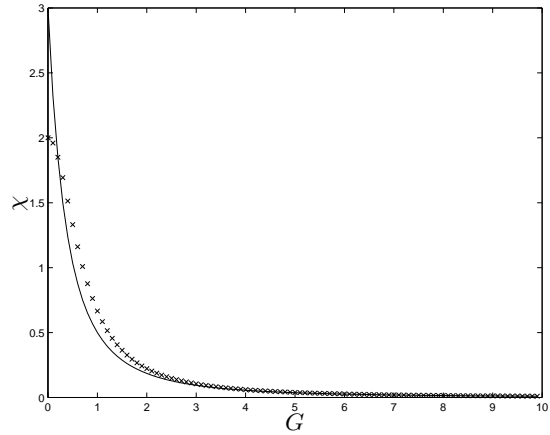}
\caption{Approximation of $\chi(G)$ for the Lorentzian distribution.  The
numerically calculated $\chi(G)$ (solid line) is compared to  $
\chi=(G^2+0.5)^{-1}$ (crosses).}
\label{fig:comp1chi}
\end{figure}

\begin{figure}[htb]
\psfrag{theta}[][]{$\theta$}
\psfrag{g}[][]{$\gamma/k_\parallel v_{Ti\parallel}$}
\psfrag{gamma}[][]{$\gamma/kv_{Ti\parallel}$}
\includegraphics[width=\mywidth,keepaspectratio]{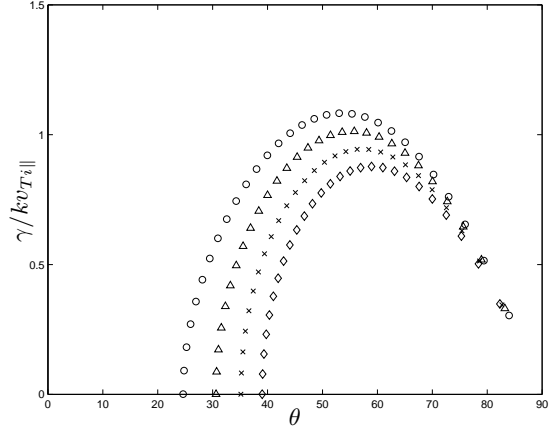}
\caption{Growth rates for the mirror instability in the
case 
of $\beta_{i\parallel}=\beta_{e\parallel}=0.1$, 
$\beta_{i\perp}=\beta_{e\perp}=0.5$, calculated with the
approximation $\bar{\chi}_i=1/(G^2+v_m^2)$, for several $v_m^2=2$
(diamonds), 1.5 (crosses), 1 (triangles), and 0.5 (circles).  The
electrons are massless Maxwellian. }
\label{fig:vapprox}
\end{figure}

The previous analysis gives a clue to the treatment of the instability
in the range of maximum growth rates, where $G\gtrsim 1$.  Let us
assume that the distribution function is such that $v_\parallel
f(v_\parallel)$ has a sharp maximum at some $v_m\sim v_{T\parallel}$. 
An example of a distribution of this kind is the Maxwellian
$f_i=(1/\sqrt{2\pi}v_{Ti\parallel})
\exp(-v_\parallel^2/2v_{Ti\parallel}^2)$ for which there  was no
good approximation for $\bar{\chi}$ in the range $|Z|\sim 1$ so far. 
For the aperiodic mirror instability with $Z=iG$, $G>0$, one has
\begin{equation}
\bar{\chi}=\int \frac{1}{iG-v_\parallel}\parvl{f} dv_\parallel=
-\int\frac{1}{G^2+v_\parallel^2} v_\parallel \parvl{f} dv_\parallel. 
\end{equation}
For $v_m\sim 1\lesssim G$ ($v_m$ is normalized on $v_{T\parallel}$) the
function $(G^2+v_\parallel^2)^{-1}$ varies slowly in the vicinity of
the maximum of $v_\parallel
(\partial f/\partial v_\parallel)$, so that one may approximate
\begin{equation}
\bar{\chi}=-\frac{1}{G^2+v_m^2} \int v_\parallel \parvl{f} dv_\parallel =
\frac{1}{G^2+v_m^2}. \label{eq:approx}
\end{equation}
Figure~\ref{fig:compchi} shows the comparison of the numerically
found $\bar{\chi}$ for the Maxwellian distribution ($v_m^2=2$) and $Z=iG, G>0$
with the approximation \eqref{eq:approx}.  The approximation proves to be
very good for $G\geq 1$ and is only by the factor 2 smaller at
$G\rightarrow 0$.  Figure~\ref{fig:comp1chi} shows similar comparison for
Lorentzian.    Now the maximum growth rate
can be obtained by substituting $\bar{\chi}_i=1/(G^2+v_{mi}^2)$ in
\eqref{eq:second}.  If $G$ is expected to be high, so that
$G^2R\mu\sim 1$, as it occurs for the Lorentzian $e-i$ distributions
in Figure~\ref{fig:threeown}, the electron inertia should be also
taken into account by substituting $\bar{\chi}_e=1/(G^2R\mu
+v_{me}^2)$.  If, however, the growth rates are relatively modest (as
in other cases studied in the present paper), the electrons still respond
adiabatically and $\bar{\chi}_e=d_e$.  In the last case
\eqref{eq:second} turns into a third order equation with respect to
$G^2$.  Finding the maximum growth rate from this equation is a
technical problem.  We shall stop for a while at the physical sense of
the above approximation.  The dependence of the maximum growth rate on
$v_m$ indicates that the particles with high velocities $v\sim
v_{T\parallel}$ are taking part in the process.  This is related to
the dynamic redistribution (closely related to the dynamic Debye
screening): if the potential changes quickly the low velocity
particles do not have enough time to change their position and leave
the field enhancements.  This redistribution is described by the same
Eq.~\eqref{eq:deltan} but now $\gamma/k_\parallel v_{T\parallel}\sim
1$.  High velocity particles can leave these regions and reduce the
kinetic pressure response but their contribution rapidly decreases
with the velocity since their number decreases.  The increase of
redistribution efficiency and the decrease of the number of screeners
with the velocity increase finds its manifestation in that the main
contribution belongs to the particles at the maximum of $v_\parallel
f(v_\parallel)$.  Since the redistribution plays the destabilizing
role, it can be expected that the smaller is $v_m$ the higher is the
growth rate.  This can be seen already from Figure~\ref{fig:fourg1}
where the growth rate for Lorentzian ions, $v_m^2=0.5$, is larger than
the growth rate for the Maxwellian, $v_m^2=2$ (with the same massless
Maxwellian electrons).  Figure~\ref{fig:vapprox} shows the comparison
of the growth rates obtained with the proposed approximation for
several $v_m^2=2$ (diamonds), 1.5 (crosses), 1 (triangles), 0.5
(circles), and massless Maxwellian electrons.  The parameters chosen
are $\beta_{i\perp}=\beta_{e\perp}=0.5$,
$\beta_{i\parallel}=\beta_{e\parallel}$.  As expected the decrease of
$v_m$ results in the increase of the maximum growth rate.

\begin{figure}[htb]
\psfrag{theta}[][]{$\theta$}
\psfrag{g}[][]{$\gamma/k_\parallel v_{Ti\parallel}$}
\psfrag{gamma}[][]{$\gamma/kv_{Ti\parallel}$}
\includegraphics[width=\mywidth,keepaspectratio]{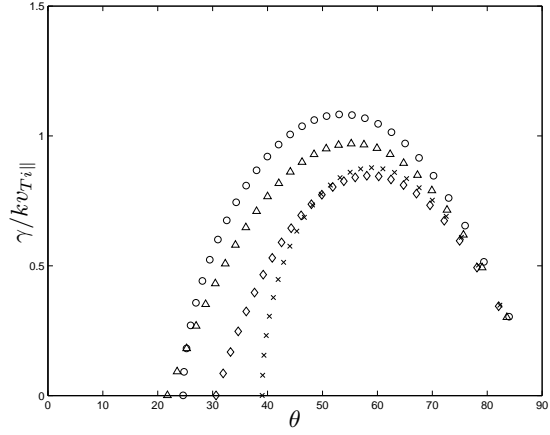}
\caption{Comparison of the growth rates for the mirror instability in the
case of $\beta_{i\parallel}=\beta_{e\parallel}=0.1$,
$\beta_{i\perp}=\beta_{e\perp}=0.5$, calculated directly and with the
approximation $\bar{\chi}_i=1/(G^2+v_m^2)$, for Maxwellian (diamonds
and crosses, respectively) and Lorentzian (triangles and circles,
respectively). The
electrons are massless Maxwellian.}
\label{fig:vapproxcomp}
\end{figure}

Finally, Figure~\ref{fig:vapproxcomp} shows the comparison of the
growth rates obtained directly and with the above approximation for
Maxwellian (diamonds and crosses) and Lorentzian (triangles and
circles), for the same parameter set.  The agreement is quite
satisfactory.

\section{Conclusions}\label{sec:concl}
We have derived the most general dispersion relation for
longwavelength modes in hot plasmas.  We have derived the general
mirror instability condition for arbitrary ion and electron
distributions and growth rate of the instability near the threshold. 
The instability threshold depends not only on the plasma species $\beta$
but also on another integral characteristic of the distribution function
$d=\int v_\parallel^{-1}(\partial f/\partial v_\parallel)
dv_\parallel$ for both species.  Larger $d$ corresponds to smaller
Debye length.  Smaller Debye length, in turn, corresponds to stronger
response of the density to the perturbations of the potential, which
allows stronger density depletions in the regions of the magnetic
field enhancements.  Therefore, the kinetic pressure response to the
magnetic pressure buildup weakens.  Hence, the larger is $d$ the lower is
instability threshold.  The near-the-threshold growth rate is
inversely proportional to $\partial f/\partial \mathcal{E}$, where
$\mathcal{E}=v_\parallel^2/2$ is the parallel energy.

The mirror instability is always aperiodic and $(\gamma/k_\parallel
v_{Ti\parallel})_{\text{max}}\sim 1$ (and sometimes substantially
greater).  Maximum growth rates are normally determined by $v_{mi}$
such that $v_\parallel\partial f_i/\partial v_\parallel$ has a sharp
maximum in $v_\parallel=v_{mi}$, and $d_e$ (if the instability is very
strong $v_{me}$ takes the place of $d_e$).  This is related to the
dynamic redistribution in which the thermal particles participate.  Growth
rates are higher for distributions with tails and lower for compact
distributions (those, for which $f=0$ if $|v_\parallel|>v_0$, where
$v_0$ is some upper limit).  For noncompact distributions the maximum
growth rate is larger for smaller $v_m$, which corresponds to the
weaker dynamic screening of the parallel electric field.  For the
distributions analyzed in this paper the behavior of $d$ and $v_m$
correlates ($d$ increases when $v_m$ decreases) since all these are
single-parameter distributions.  For more general distributions the
behavior of $d_e$ and $v_m$ may be uncorrelated.  It is also worth
noting that it not, in general, any specific group of particles which
are responsible for the instability development.  Compare, for
example, two similar distributions (velocity normalized
on the thermal velocity $v_{T\parallel}$):
$f_1=(2/\pi)(1+v_\parallel^2)^{-2}$ with $d=3$ and $v_m^2=0.5$, and
$f_2=(\sqrt{2}/\pi)(1+v_\parallel^4)^{-1}$ with $d=1$ and $v_m^2=1$. 
While the behavior of the two is similar for $v_\parallel=0$ and
$v_\parallel\rightarrow \infty$ (the only difference is the factor
$\sqrt{2}$), the first one is expected to be more unstable because of
the three times stronger Debye screening.  At the same time the
behavior of the second distribution near the threshold should be close
to that of the Maxwellian, $d=1$, despite the very different
suprathermal tails and $(df/d\mathcal{E})|_{v_\parallel=0}$.

We have also proposed a useful approximation for the dielectric
function in the range $G/k_\parallel v_{Ti\parallel}\gtrsim 1$ for
distributions with sharp maxima of $v_\parallel(\partial f/\partial v_\parallel) $
(Maxwellian as one of such distributions).  This approximation
proves to be quite satisfactory for Maxwellian type distributions
and allows to study analytically the instability behavior in the maximum
growth rate range.

\begin{acknowledgments}
Figures are made using Matlab. 
\end{acknowledgments}

\appendix
\section{General expressions}\label{sec:genexp}

We start with the general expression for the dielectric tensor in the 
following form:
\begin{equation}
\epsilon_{ij}=\delta_{ij} +\sum\lambda_{ij}, \label{eq:gengen}
\end{equation}
where the summation is on the species and 
\begin{equation}
\lambda_{ij}=-\frac{\omega_p^2}{\omega^2} \delta_{ij}+\eta_{ij}.
\label{eq:gen1}
\end{equation}
The expression for $\eta_{ij}$ is well-known (see, e.g., \textcite{Ha75}):
\begin{equation}
\eta_{ij}=-\sum_n\frac{\omega_p^2}{\omega^2} \int v_\perp dv_\perp 
dv_\parallel \left(\frac{n\Omega}{v_\perp}\frac{\partial 
f_0}{\partial 
v_\perp}+k_\parallel \frac{\partial f_0}{\partial v_\parallel} \right)
\frac{\Pi_{ij}}{n\Omega -\zeta},
\label{eq:general}
\end{equation}
where $\zeta=\omega-k_\parallel v_\parallel$, and 
\begin{equation}
\label{eq:pij}
\Pi_{ij}=\begin{pmatrix}
(n^2\Omega^2/k_\perp^2)J_n^2 & i(v_\perp n\Omega/k_\perp) 
J_nJ'_n & (v_\parallel n\Omega/k_\perp) J_n^2\\
-i(v_\perp n\Omega/k_\perp) 
J_nJ'_n & v_\perp^2 {J'_n}^2 & -iv_\perp v_\parallel J_nJ'_n \\
(v_\parallel n\Omega/k_\perp) J_n^2 & iv_\perp v_\parallel 
J_nJ'_n & v_\parallel^2J_n^2
\end{pmatrix}.
\end{equation}
Here $J_n=J_n(x)$, $x=k_\perp\rho=k_\perp v_\perp/\Omega$, and 
$J'_n=dJ_n/dx$.

For the analysis in the low-frequency range $\omega/\Omega\ll 1$ let
us write
\begin{equation}
\eta_{ij}= 
\eta_{ij}^{(0)} +\eta_{ij}^{(n\ne0)}, \label{eq:decomp}
\end{equation}
and expand 
\[
\frac{1}{n\Omega-\zeta}=\frac{1}{n\Omega}
\left(1+\frac{\zeta}{n\Omega} +
\frac{\zeta^2}{n^2\Omega^2} +\cdots\right).
\]

Let also $f_0=f_1(v_\perp)f_2(v_\parallel^2)$, and denote 
$\aver{\ldots}=\int (\ldots) f dv_j$, where $j=\perp,\parallel$.

One has
\begin{equation}\label{eq:eta0}
\eta_{ij}^{(0)}=\wwp\int v_\perp dv_\perp 
dv_\parallel \frac{k_\parallel}{\zeta}\frac{\partial f_0}{\partial 
v_\parallel}\times \begin{pmatrix} 0 & 0 & \\
0 & v_\perp^2{J'_0}^2 & -iv_\perp v_\parallel J_0J'_0\\
0 & iv_\perp v_\parallel J_0J'_0 & v_\parallel^2J_0^2
\end{pmatrix}
\end{equation}
and
\begin{equation}\label{eq:etan}
\begin{split}
& \eta_{ij}^{(n\ne0)}=-\sum\wwp\int v_\perp dv_\perp 
dv_\parallel \left(\frac{1}{v_\perp} \frac{\partial 
f_0}{\partial v_\perp} + \frac{k_\parallel}{n\Omega}
\frac{\partial f_0}{\partial 
v_\parallel}\right) \\
& \times \left(1+\frac{\zeta}{n\Omega} + 
\frac{\zeta^2}{n^2\Omega^2}\right) \Pi_{ij}.
\end{split}
\end{equation}

Now, up to  $\Omega^{-2}$ one obtains 
\begin{align}
& \eta_{11}^{(n\ne0)}=-\sum_n\wwp 
\left[\aver{J_n^2\frac{\partial}{\partial v_\perp}} 
\frac{n^2\Omega^2+\omega^2 
+ k_\parallel^2 \aver{v_\parallel^2}}{k_\perp^2}
+\frac{k_\parallel^2}{k_\perp^2}\aver{v_\perp J_n^2} \right], 
\label{eq:eta11ne0}\\
& \eta_{12}^{(n\ne0)}=-i\sum_n\wwp \frac{\omega}{k_\perp} 
\aver{v_\perp J_nJ'_n\frac{\partial}{\partial v_\perp}},
\label{eq:eta12ne0}\\
& \eta_{13}^{(n\ne0)}=\sum_n\wwp \frac{k_\parallel}{k_\perp} 
\left[\aver{v_\perp J_n^2} +\aver{v_\parallel^2} \aver{J_n^2\parvp} 
\right],\label{eq:eta13ne0}\\
&\eta_{22}^{(n\ne0)}=-\sum_n\wwp \left[\aver{v_\perp^2 {J'_n}^2 
\frac{\partial }{\partial v_\perp}} \left(1 + \frac{\omega^2+ 
k_\parallel^2\aver{v_\parallel^2}}{n^2\Omega^2}\right)
+\frac{k_\parallel^2}{n^2\Omega^2}\aver{v_\perp^3{J'_n}^2}\right], 
\label{eq:eta22ne0}\\
& \eta_{23}^{(n\ne0)}=-i\sum_n\frac{\omega_p^2k_\parallel}{\omega 
n^2\Omega^2}\left[\aver{v_\perp^2J_nJ'_n} + 2\aver{v_\parallel^2} 
\aver{v_\perp J_nJ'_n \parvp{}}\right], \label{eq:eta23ne0}\\
& \eta_{33}^{(n\ne0)}=-\sum_n\wwp \aver{v_\parallel^2} \aver{J_n^2 
\frac{\partial }{\partial v_\perp}} ,\label{eq:eta33ne0}
\end{align}
and
\begin{align}
& \eta_{22}^{(0)}= \wwp k_\parallel \aver{v_\perp^3{J'_0}^2} 
\chi,\label{eq:eta220}\\
&  \eta_{23}^{(0)}=-i \frac{\omega_p^2}{\omega}
\aver{v_\perp^2J_0J'_0} \chi, \label{eq:eta230}\\
& \eta_{33}^{(0)}= \wwp  \aver{v_\perp J_0^2} 
\left(1+\frac{\omega^2}{k_\parallel}\chi\right).\label{eq:eta330}
\end{align}
where 
\begin{equation}
\chi=\aver{\frac{1}{\zeta}\parvl{}}.\label{eq:chi}
\end{equation}
Using in Eqs.\eqref{eq:eta11ne0}-\eqref{eq:eta33ne0} the following
summation 
 rules
\begin{align}
& \sum_{n\ne 0} J_n^2=1-J_0^2,\\
& \sum_{n\ne 0} 
n^2J_n^2=\frac{x^2}{2}= \frac{k_\perp^2v_\perp^2}{2\Omega^2},\\
& \sum_{n\ne0}J_nJ'_n=-J_0J'_0,\\
& \sum_{n\ne0}{J'_n}^2=\frac{1}{2}-{J'_0}^2,
\end{align}
one obtains eventually the following general expression for 
$\lambda_{ij}$ in the limit of $\omega,k_\parallel v_\parallel \ll 
\Omega$ when expanded up to the second order in $\zeta/\Omega$:
\begin{align}
& \lambda_{11}= \frac{\omega_p^2}{k_\perp^2} \left(1 + 
\frac{k_\parallel^2 \aver{v_\parallel^2}}{\omega^2} \right) 
\aver{J_0^2 \parvp{}} -\wwp \frac{k_\parallel^2}{k_\perp^2} 
\aver{v_\perp(1-J_0^2)}, \label{eq:la11}\\
& \lambda_{12} = i\frac{\omega_p^2}{\omega k_\perp} \aver{v_\perp 
J_0J'_0 \parvp{}}, \label{eq:la12}\\
& \lambda_{13}=\wwp \frac{k_\parallel}{k_\perp} 
\left[\aver{v_\perp(1 - J_0^2)} - \aver{v_\parallel^2} \aver{J_0^2 
\parvp{}}\right],\label{eq:la13}\\
& \lambda_{22}= \wwp \aver{v_\perp^2{J'_0}^2 \parvp{}} 
-\wwp \sum_n\left[\frac{\omega^2 + k_\parallel^2\aver{v_\parallel^2}} 
{n^2\Omega^2} \aver{v_\perp^2 {J'_n}^2 \parvp{}} \right. 
\label{eq:la22}\\
&\left.+\frac{k_\parallel^2}{n^2\Omega^2} \aver{v_\perp^3 
{J'_n}^2}\right] 
 +\wwp k_\parallel \aver{v_\perp^3{J'_0}^2} 
\chi,\notag\\
& \lambda_{23}=
-i\sum_n\frac{\omega_p^2k_\parallel}{\omega 
n^2\Omega^2}\left[\aver{v_\perp^2J_nJ'_n} + 2\aver{v_\parallel^2} 
\aver{v_\perp J_nJ'_n \parvp{}}\right] 
-i \frac{\omega_p^2}{\omega}
\aver{v_\perp^2J_0J'_0} \chi, \label{eq:la23}\\
& \lambda_{33}= -\wwp \left[1-\aver{v_\parallel^2} \aver{J_0^2 
\parvp{}} \right]+ 
\wwp  \aver{v_\perp J_0^2} 
\left(1+\frac{\omega^2}{k_\parallel}\chi\right).\label{eq:la33}
\end{align}

It is possible to get rid of the series in
\eqref{eq:la22}--\eqref{eq:la33} using \cite{Prud}
\begin{align}
& \Phi=\sum_{n\ne 0}\frac{J_n^2}{n^2}=\frac{4}{\pi} \int_0^{\pi/2} t^2
J_0(2x\cos t) dt- \frac{\pi^2}{6}J_0^2, \label{eq:philarge}\\
& \Psi=\sum_{n\ne 0}\frac{J_nJ'_n}{n^2}=\tfrac{1}{2}\frac{d\Phi}{dx},
\label{eq:psilarge}\\
& \Upsilon=\sum_{n\ne 0}\frac{(J'_n)^2}{n^2}=
\tfrac{1}{2}\left[\frac{d^2\Phi}{dx^2} +\frac{1}{x} \frac{d\Phi}{dx}
+2\Phi -\frac{2}{x}(1-J_0^2)\right]. \label{eq:ularge}
\end{align}
This may be useful for calculations in the regime $k_\perp
v_\perp/\Omega\sim 1$.

The general dispersion relation is obtained from the determinant
$\det\vert D\vert=0$, where $D_{ij}=N^2\delta_{ij} - N_iN_j 
-\epsilon_{ij}$, that is, 
\begin{align}
& D_{11}=N_\parallel^2 -1 -\sum \lambda_{11}, \label{eq:D11}\\
& D_{12}=-\sum \lambda_{12}, \label{eq:D12}\\
& D_{13}= -N_\parallel N_\perp -\sum\lambda_{13}, \label{eq:D13}\\
& D_{22}=N^2 -1-\sum \lambda_{22}, \label{eq:D22}\\
& D_{23}=-\sum \lambda_{23}, \label{eq:D23}\\
& D_{33}=N_\perp^2 -1-\sum \lambda_{33}. \label{eq:D33}
\end{align}
The polarization should be found from the equations
\begin{equation}
D_{ij}E_j=0. \label{eq:pol}
\end{equation}
Eq.~\eqref{eq:pol} provides the ratio of the electric field 
components. In order to translate that into the magnetic polarization 
one has to use the relation 
$\mathbf{B}=\mathbf{k}\times\mathbf{E}/\omega$.
In order to find the density perturbations one has to use the current 
conservation as follows
\begin{equation}
\delta \rho = \mathbf{k}\cdot\mathbf{j}/\omega, \label{eq:cc}
\end{equation}
where 
\begin{equation}
j_i=-i\frac{\omega}{4\pi}\lambda_{ij}E_j, \label{eq:sigma}
\end{equation}
so that one has eventually
\begin{equation}
\delta \rho=-\frac{i}{4\pi} k_i\lambda_{ij}E_j.\label{eq:dens}
\end{equation}

Further simplifications are possible in the longwavelength limit.

\section{Longwavelength approximation}\label{sec:long}
In this appendix we  provide general expressions for the dielectric 
tensor in the longwavelength limit $k_\perp v_\perp/\Omega\ll 1$, 
where
$J_{\pm 1}=\pm k_\perp v_\perp/2\Omega$,
$J_0=1-k_\perp^2v_\perp^2/2\Omega^2$, and higher order Bessel
functions may be neglected.  In this limit one has
\begin{align}
& \lambda_{11}=\frac{\omega_p^2}{\Omega^2} 
+\tfrac{1}{2}N_\parallel^2(\beta_\parallel-\beta_\perp), 
\label{eq:la11lw}\\
& \lambda_{12}=i \frac{\omega_p^2}{\omega\Omega}, \label{eq:la12lw}\\
& \lambda_{13}= -\tfrac{1}{2}
N_\parallel N_\perp (\beta_\parallel-\beta_\perp), 
\label{eq:la13lw}\\
& \lambda_{22}=\frac{\omega_p^2}{\Omega^2} 
+\tfrac{1}{2}N_\parallel^2(\beta_\parallel-\beta_\perp)
\label{eq:la22lw}\\
& -N_\perp^2\beta_\perp +\frac{N_\perp^2\beta_\perp}{4} 
\frac{\aver{v_\perp^4}}{\aver{v_\perp^2}} \chi], \notag\\
& \lambda_{23}=i \frac{\beta_\perp \tan\theta\Omega}{2\omega}c^2\chi, 
\label{eq:la23lw}\\
& \lambda_{33}= \tfrac{1}{2} 
N_\perp^2(\beta_\parallel-\beta_\perp) + (
\frac{\omega_p^2}{k_\parallel^2} -\frac{\beta_\perp \tan^2\theta 
c^2}{2})\chi, 
\label{eq:la33lw}
\end{align}
where $N=kc/\omega$, $N_\perp=k_\perp c/\omega=N\sin\theta$, 
$N_\parallel=lk_\parallel c/\omega=N\cos\theta$, 
$\beta_\parallel=2\omega_p^2 \aver{v_\parallel^2}/c^2\Omega^2$, 
$\beta_\perp=\omega_p^2 \aver{v_\perp^2}/c^2\Omega^2$, 
and $\aver{\ldots}$ denotes usual averaging over the distribution.
Here also $\zeta=u-v_\parallel$, where 
$u=\omega/k_\parallel$.  The last term in~\eqref{eq:la33lw} is given
for completeness.  In the limit used in this paper,
$\omega/\Omega\rightarrow 0$ and $\omega/k$ finite,   it
should be neglected.  Throughout the paper we also assume
$\omega_{pi}\gg \Omega_i$.

\section{Aperiodic nature of the mirror instability}
\label{sec:aperiodic}

In order to show that the mirror instability is aperiodic we analyze
the behavior of the roots of \eqref{eq:second} when the
$\beta$ parameters are changed.  In the waterbag case the transition from
the stable to the unstable regimes  occurs when $W=0, G=0$ and for the mode
whose phase velocity is less than the highest particle velocity,
$\real Z<v_{\parallel,\text{max}}$ (``subparticle'' mode), that is, in the
resonant region.  In the general case, where Landau damping is
nonzero, in the resonant range every propagating wave having $W\ne 0$
has also nonzero damping rate $G<0$ (we assume that there are no other
kinetic instabilities in the mirror-stable region).  By continuously
changing the plasma parameters (e.g., the anisotropy ratio
$\beta_\perp/\beta_\parallel$) we can bring the system into the
unstable regime.  Assuming continuous dependence of $W$ and $G$ on the
plasma parameters we see that it is impossible for the ``subparticle''
mode with $W\ne0$ to transform into the unstable mode, since $G<0$ and
cannot be made positive continuously.  Thus, the only way to do that
is to go through $W=0$, $G=0$.

Let us now consider the vicinity of the transition to the instability,
$|Z|\ll 1$.  In the most general way, expanding $\bar{\chi}$ in powers
of $Z$ one gets:
\begin{equation}\label{eq:powers}
\begin{split}
&\bar{\chi}=\int \frac{1}{Z-v_\parallel}\parvp{f} dv_\parallel \\
& = - \frac{1}{v_\parallel}\parvp{f}dv_\parallel +Z \int 
\frac{1}{Z-v_\parallel}\frac{1}{v_\parallel}\parvp{f} dv_\parallel \\
& \rightarrow d+i\kappa Z, 
\end{split}
\end{equation}
provided $(\partial f/\partial v_\parallel)|_{v_\parallel=0}\ne 0$. 
The quantities $d$ and $\kappa$ are defined in~\eqref{eq:chith}.
It is easy to see that \eqref{eq:second} is the first order equation 
for $iZ$ (with real coefficients) in the lowest order on $|Z|\ll 1$, 
which means that there is a simple (one and only one) aperiodic root
 in the vicinity of $Z=0$.  Such aperiodic solutions cannot be
converted into non-aperiodic ones by continuous change of the plasma
parameters, for the same reason as above.  Therefore, the unstable
solutions must be aperiodic.

The function
$\Psi(Z)$, defined in~\eqref{eq:second}, is an analytical function of
$Z=W+iG$ and a continuous function of its parameters $\beta$ and
$\theta$.  Let us consider how $Z$ moves from the lower half-plane
(stable regime) to the upper half-plane (unstable regime) with the
change of $\beta$ and $\theta=\text{const}$.  The transition to
instability occurs, in general, in the vicinity of $Z=0$ where
$\Psi(Z)=\Psi(0)+(d\Psi/dZ)|_{Z=0}Z=A+BZ$ (see
sec.~\ref{sec:threshold}).  In the transition point $A=0$.  Using
\eqref{eq:G}--\eqref{eq:B} it is easy to show that in the transition
point $B>0$ (provided $\kappa>0$, this condition being violated if
$\partial f/\partial \mathcal{E}>0$ at $v_\parallel=0$, corresponding
to the regime of two-hump instability), so that in the vicinity of the
transition point $A>0$ corresponds to the stable regime, while $A<0$
corresponds to the instability.  Because of the continuity, in the
whole instability range $A<0$.

Let us show now that \eqref{eq:second} always has a solution $Z=iG$,
$G>0$ in the unstable range.  Indeed, $\Psi(0)<0$ as is shown above. 
On the other hand, if $G\rightarrow \infty$ one has $\chi\rightarrow
1/G^2$ and $\Psi(\infty)>0$.  This means that there exists $G>0$ such
that $\Psi(G)=0$. 

In the absence of kinetic instabilities, in the stable regime all roots of
\eqref{eq:second} with nonzero $W$ are either in the lower half-plane
(Landau damping or nonpropagation) or at the real axis (if $(\partial
f/\partial v_\parallel)|_{v_\parallel=W}=0)$.  In the first case no
root can cross the real axis except at $W=0$, when the parameters are
changed continuously to bring the system in the unstable regime.  As
can be seen from \eqref{eq:G}--\eqref{eq:B} there is only one root
crossing the real axis at this point, provided 
$\partial f/\partial \mathcal{E}<0$.
Therefore, there is only
one root in the upper half-plane and it is purely imaginary.

If $\partial f/\partial \mathcal{E}=0$ there are two or more
roots in the vicinity of $Z=0$ (depending on the behavior of $f$) but only
one is positive, $G>0$.  Since in this case the analytical
continuation through $Z=0$ into the lower half-plane is
straightforward (no pole at $v_\parallel=0$) other roots correspond to
damping solutions, and there is again only one root in the upper
half-plane.

Finally, let us consider the case where there are roots with $G=0$ and
$W=W_0\ne 0$.  Such situation can occur when $(\partial f/\partial
v_\parallel)=0$ in isolated points or in an interval (as for the
compact waterbag and hard-bell).  In the first case the imaginary part
of $Z$ is negative for $W$ close to $W_0$, so that the continuous
change of parameters does not bring the root to the upper half-plane. 

In the second case the continuous change of parameters leaves the root
on the real axis until it enters the range where $(\partial f/\partial
v_\parallel)\ne0$ or $W=0$.

\newpage

\end{document}